\documentclass[aps,twocolumn,preprintnumbers,groupedaddress,nofootinbib,pra]{revtex4-2}
\pdfoutput=1

\usepackage[usenames,dvipsnames,table]{xcolor}
\usepackage{graphicx,amsmath,amssymb,amsthm,multirow,array,bm,bbm,esint}
\usepackage{enumitem}
\usepackage[mathscr]{eucal}
\usepackage[bbgreekl]{mathbbol}
\usepackage{epsf,amsfonts}
\usepackage{physics}
\usepackage{dsfont, mathtools}
\usepackage{slashed}
\usepackage{hyperref}
\usepackage{mathdots}
\usepackage{upgreek}
\usepackage{comment}

\usepackage[normalem]{ulem}

\hypersetup{
    pdfstartview={FitH},    
    pdftitle={Holographic black hole formation},    
    colorlinks=true,       
    linkcolor=blue,          
    citecolor=red,        
    filecolor=magenta,      
    urlcolor=blue           
}

\usepackage{tikz}

\newcommand{\hypergeometricF}[5]{ ~_{#1}F_{#2} \left[ \left.\begin{matrix} #3 \\ #4 \end{matrix}\right| #5 \right]}

\makeatletter 
\renewcommand\onecolumngrid{
\do@columngrid{one}{\@ne}%
\def\set@footnotewidth{\onecolumngrid}
\def\footnoterule{\kern-6pt\hrule width 1.5in\kern6pt}%
}
\makeatother    

\begin{document}

\title{Holographic Black Hole Formation\\ and Scrambling in Time-Ordered Correlators}

\author{Pratyusha Chowdhury}
\affiliation{School of Physics and Astronomy \& STAG Research Centre, University of Southampton, SO17 1BJ, Southampton, United Kingdom}
\author{Felix M.\ Haehl, Adri\'{a}n S\'{a}nchez-Garrido}
\affiliation{School of Mathematical Sciences \& STAG Research Centre, University of Southampton, SO17 1BJ, Southampton, United Kingdom}
\author{Ying Zhao}
\affiliation{MIT Center for Theoretical Physics -- a Leinweber Institute, Massachusetts Institute of Technology, Cambridge, MA 02139, USA}

\begin{abstract}
We describe a holographic mechanism for black hole formation via the collision of two shock waves in three-dimensional anti-de Sitter spacetime. In the dual conformal field theory (CFT), a two-shock-wave state corresponds to the insertion of two boosted ``precursor’’ operators in complementary Rindler patches. Their operator product expansion is initially described by a universal mean-field spectrum of exchanged states, which is dominated by operator dimensions that grow exponentially in the boost parameter. We propose their mean value as diagnosing the mass of the collision product in the bulk. It crosses the CFT heavy state  threshold after two scrambling times, in accordance with expectations about black hole formation in general relativity. Our analysis also allows us to identify the scrambling characteristics usually associated with out-of-time-order correlation functions, using only the internal composition of thermal in-time-order correlators.
\end{abstract}

\maketitle

\section{Introduction}
\label{sec:intro}

While the holographic duality has led to profound insights into quantum gravity, deep puzzles about black hole dynamics remain. To address these microscopically, it is of paramount importance to understand the process of black hole formation in the language of the dual conformal field theory (CFT). It has long been known that black holes in three-dimensional anti de Sitter (AdS) spacetime \cite{BTZ1,BTZ2} can be formed in two ways: $(i)$ by gravitational collapse of a dust shell \cite{Ross:1992ba,Danielsson:1999fa,Danielsson:1999zt,Giddings:1999zu,Giddings:2001ii}, $(ii)$ by collision of shock waves \cite{Gott-PRL,Matschull:1998rv,Holst:1999tc,Polchinski:1999yd,Balasubramanian:1999zv,Horowitz:1999gf,Lindgren:2015fum}. While a holographic understanding in terms of the conformal operator product expansion (OPE) was initiated for the first option in \cite{Anous:2016kss}, such a perspective has long remained a challenge for the second.

\begin{figure}
\includegraphics[width=.8\columnwidth]{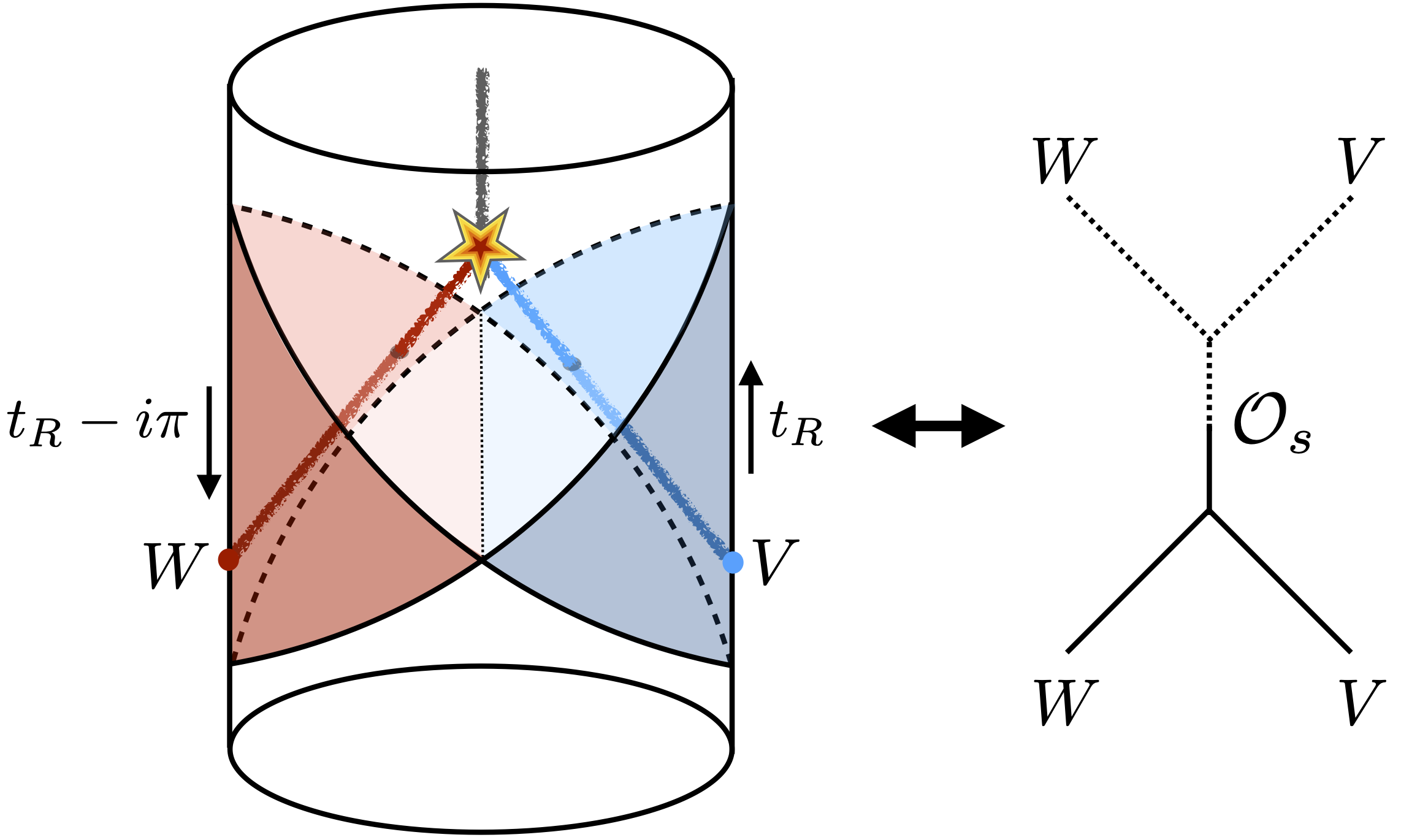}
\caption{Kinematic setup: each shock wave originates in one AdS$_3$ Rindler patch. Information about the collision product is contained in the distribution of exchanged operators in the cross-channel OPE, which depends strongly on the boost.}
\label{fig:setup}
\end{figure}

We study this problem in the cleanest holographic setup: the collision of two gravitational shock waves, focused onto each other in an empty global AdS$_3$ spacetime. We prepare the two-shock ``microstate'' using CFT primary operators boosted back in time with the Rindler Hamiltonian (so-called precursor operators). For a single operator, Rindler time evolution leads to an exponential spreading of the operator within its conformal family \cite{Caputa:2021sib,Caputa:2021ori}. In gravity, this corresponds to an increasingly energetic, nearly null shock wave. However, this kinematical effect by itself cannot lead to black hole formation \cite{Keller:2014xba}. In order to set up black hole formation, we require two operators to interact {\it dynamically} via the OPE, see Fig.\  \ref{fig:setup}. In this Letter we quantify how universal dynamical input leads to an exponential spreading across the space of possible exchanged operators ${\cal O}_s$, allowing us to make detailed predictions about the collision product.

The connection between operator growth, information scrambling, and the out-of-time-order correlator (OTOC) is well known \cite{Sekino:2008he,Maldacena:2015waa,Shenker:2013pqa}. In this work we add a novel and surprising object to this list of quantum chaos diagnostics: the {\it in-time-order} four-point correlation function (TOC) of pairwise identical boosted operators. This arises naturally as the self-overlap of the two-shock-wave state. Naively, the TOC's time dependence is trivial; however, its decomposition into irreducible components turns out to give access to time dependent scrambling behavior normally associated with the OTOC.

\section{Setup}
\label{sec:setup}

\subsection{Warmup: one shock wave}

Consider a Rindler patch of global AdS$_3$ with Rindler coordinates $(t_R,x)$. A gravitational shock wave can be prepared in the dual CFT by acting with a primary operator $W$ with conformal weight $1\ll \Delta_w \ll c$:
\begin{equation}
\label{eq:PhiDef}
    |\Phi_{W_L}\rangle = W(-t_w-i\pi+i\delta,x_w) |0\rangle \,,
\end{equation}
where $t_w<0$ and $|0\rangle$ is the global vacuum state, which a Rindler observer experiences with an inverse temperature $\beta = 2\pi L_\text{AdS}$. In the following we set the AdS radius $L_\text{AdS}=1$. Note that the argument of the operator is expressed in Rindler coordinates, and, therefore, the shift by $-i\pi$ means that the operator $W$ is inserted in the left Rindler patch (indicated by subscript `L'), where bulk time is directed in the opposite direction, see Fig.\  \ref{fig:setup}. The explicit transformation between plane and Rindler coordinates can be found in the Supplemental Material (SM) Sec.\ \ref{app:CrossRatios}. The small imaginary shift $i\delta$ is required to produce a localized shock wave with finite energy \cite{Afkhami-Jeddi:2017rmx}.

We increase $-t_w$ starting from $0$. This corresponds to evolving the operator into the past with the CFT boost generator $i\partial_{t_R}$. For large $-t_w$, the energy of the excitation localizes along null directions. For example, the expectation value of the light-cone stress-energy tensor is given by a localized shock \cite{Afkhami-Jeddi:2017rmx}:
\begin{equation}
 \frac{\langle \Phi_{W_L} | T_{++}(t-i\pi,x) | \Phi_{W_L} \rangle}{\langle \Phi_{W_L} | \Phi_{W_L} \rangle} \sim \frac{h_w}{\sin(\delta)} \, \delta((t_w+x_w)-(t+x)) 
\end{equation}
A more microscopic picture is as follows. Under Rindler time evolution, an increasing number of global conformal descendants of $W$ are populated: the primary state ``grows'' into a coherent superposition of descendants (which are eigenstates of the global Hamiltonian), centered around an increasingly high level (see \cite{Caputa:2021ori,Caputa:2021sib,Caputa:2022zsr} for a detailed analysis).
In the dual gravitational theory, the small past perturbation evolves into an almost-null shock wave whose proper energy on the $t_R=0$ slice increases exponentially in $-t_w$. Its gravitational backreaction (on other probes) becomes significant and can be described by a shock wave geometry after a scrambling time \cite{Shenker:2013pqa,Banados:1998gg}:
\begin{equation}
    -t_w \sim t_* \sim  \; \log \left(\frac{1}{\Delta_w\,G_N}  \right)\,.
\end{equation}

Note that this setup never produces a black hole in the bulk, regardless of the value of $-t_w$: the invariant rest mass of the boosted particle remains unchanged. The boosting of the operator is purely kinematical and does not excite boundary graviton degrees of freedom \cite{Keller:2014xba}. It is dependent on the choice of a reference frame and can be ``undone'' by a global isometry transformation. Similarly, in the CFT, the irreducible representation of the Virasoro algebra is invariant and remains the one labeled by $W$ for all times.

\subsection{Two shock waves}

To allow for the possibility of dynamical black hole formation, consider now a two-shock-wave state (Fig.\  \ref{fig:setup}):
\begin{equation}\label{eq:Two_shock_state}
\begin{split}
 |\Psi_{W_LV_R}\rangle  &=  W(-t_w-i\pi+i\delta,x_w) V(t_v-i\delta,x_v) |0\rangle\,,
\end{split}
\end{equation}
where $t_v, t_w <0$. Rindler time evolution increases the time difference $t\equiv-t_v-t_w >0$. This corresponds to boosting the operators in the global reference frame: in the bulk, the energy of each particle in the center of mass frame increases exponentially with $t$. Their collision initially causes a conical defect geometry to form, which can be characterized by its mass $M$ and spin $J$. This process was studied from a purely gravitational perspective in \cite{Matschull:1998rv,Holst:1999tc}. By analyzing the special geometric features of conical defects in AdS$_3$, it was found that a BTZ black hole forms when the mass exceeds the threshold value set by the extremality bound \cite{Banados:1992wn,Keller:2014xba}:
\begin{equation}
\label{eq:gott}
    M \geq |J| \,.
\end{equation}
This bound is analogous to the so-called Gott condition in flat spacetimes \cite{Gott-PRL}. It was also argued for using holographic quantum circuit models in \cite{Haehl:2023lfo}. See also \cite{Das:2017cnv,Banerjee:2024qgg} for related recent ideas.
As we show below, this condition reads as follows in terms of the parameters of the CFT precursor state:
\begin{equation}
\label{eq:gott2}
    \frac{\sqrt{\Delta_v\Delta_w}}{\sin\delta} \, e^{\frac{t-|b|}{2}} \geq \frac{c}{12} \equiv \frac{1}{8 G_N} \,, 
\end{equation}
where $b=x_v-x_w$ is the impact parameter.

We derive the threshold condition \eqref{eq:gott2} by analyzing the state $|\Psi_{W_LV_R}\rangle$ from a microscopic perspective, using conformal bootstrap tools.
Similar precursor states have been analyzed extensively in the study of quantum chaos \cite{Shenker:2013pqa,Roberts:2014isa,Shenker:2014cwa}. In particular, the scrambling time can be defined as the timescale where naive large-$N$ factorization breaks down because the OTOC $\langle \Psi_{V_LW_R} | \Psi_{W_LV_R}\rangle$ deviates significantly from $\langle V V \rangle \langle W W \rangle$. Crucially, the OTOC computes the overlap of two {\it differently ordered} states, indicated by `L' and `R' labels, referring to the Rindler patch in which operators are inserted, and its path integral representation requires a twice-folded time contour \cite{Aleiner:2016eni}. The difference between the states amplifies over time and leads to a breakdown of large-$N$ factorization after a scrambling time $t_*$.

In this work, we give a more intrinsic description of the two-shock-wave state $|\Psi_{W_LV_R}\rangle$. We want to ask: {\it How does the decomposition of the two-shock state into irreducible representations of the Virasoro algebra change over time?}\footnote{A related approach is to quantify the operator size of the two-shock state by computing a probe correlator $\langle \Psi_{W_LV_R} | {\cal O}(t_1) {\cal O}(t_2) |\Psi_{W_LV_R}\rangle$. This is again an OTOC, now requiring a thrice-folded time contour \cite{Haehl:2017pak,Haehl:2021tft}.}
Instead of OTOCs, we consider simply the self-overlap of the (unnormalized) state $|\Psi_{W_LV_R}\rangle$, which we refer to as an {\it in-time-order four-point function:}
\begin{equation}
\label{eq:FTOCdef}
    {\cal F}_\text{TOC} \equiv \frac{\langle\Psi_{W_LV_R}| \Psi_{W_LV_R}\rangle}{\langle V V\rangle \langle W W\rangle} 
    \equiv\frac{\text{tr}( W^\dagger W \rho_0^{\frac{1}{2}} V^\dagger V \rho_0^{\frac{1}{2}})}{\text{tr}( V^\dagger V\rho_0) \text{tr}( W^\dagger W\rho_0)} 
    \,,
\end{equation}
where operators in the left Rindler wedge are inserted in an analytically continued fashion halfway around the thermal circle. Indeed, $\rho_0 = \frac{1}{Z}\,e^{-2\pi H}$ is the thermal density matrix seen by a Rindler observer evolving with respect to the Rindler boost generator $H$. Real-time dependence of operators $V$ and $W$ is implicit in \eqref{eq:FTOCdef}, while imaginary time dependence is accounted for by $\rho_0^{1/2}$ factors.
This correlator is time ordered, hence the label `TOC'.
 Its value in a large-$N$ chaotic CFT is ${\cal F}_\text{TOC} \approx 1$ to a  good approximation for all times due to large-$N$ factorization.
We will not be interested in the value of ${\cal F}_\text{TOC}$, but in the details of the internal decomposition of the correlator. Perhaps surprisingly, this will unveil scrambling dynamics.
 

\subsection{Crossing equation and conformal block decomposition}

We begin with a decomposition of the state $|\Psi_{W_LV_R}\rangle$ into irreducible representations of the Virasoro algebra, i.e., Virasoro ``OPE blocks'' labeled by all possible exchanged primary operators ${\cal O}_s$ \cite{Czech:2016xec,deBoer:2016pqk,Fitzpatrick:2016mtp,Haehl:2025ehf}:
\begin{equation}
\label{eq:OPEblocks}
    |\Psi_{W_LV_R}\rangle \propto \sum_{{\cal O}_s} C_{wvs} \, |{\cal B}_{WV{\cal O}_s}(t_w,x_w;t_v,x_v)\rangle
\end{equation}
The OPE blocks are bilocal operators, furnishing an orthogonal basis of physical exchanges over which we can expand $|\Psi_{W_LV_R}\rangle$ and study its ``size'' and ``spread''.

It will be more convenient to analyze the same decomposition by computing the self-overlap, \eqref{eq:FTOCdef}. Assuming large-$N$ factorization and a gap in the spectrum of exchanged dimensions, it is clear that ${\cal F}_\text{TOC} \approx 1$. This is particularly true in CFTs with a gravity dual, where the conformal block associated with the identity operator dominates \cite{Hartman:2013mia}. We refer to this process ($W W \rightarrow \mathbb{1} \rightarrow VV$) as {\it identity dominated t-channel} exchange.\footnote{In the case of the OTOC, identity dominance is rather subtle to establish due to non-convergence of the OPE \cite{Chang:2018nzm}. In our setup, using ${\cal F}_\text{TOC}$, the OPE converges even as $z\rightarrow 1^-$. Identity dominance is thus justified as long as the spectrum of t-channel exchanges is sufficiently gapped.}

By crossing symmetry, we can equivalently decompose ${\cal F}_\text{TOC}$ into Virasoro conformal blocks in the {\it s-channel} ($W V \rightarrow {\cal O}_s \rightarrow W  V$), directly inherited from \eqref{eq:OPEblocks}:
\begin{equation}
\begin{split}
 1\approx {\cal F}_\text{TOC} &= \frac{(1-z)^{2h_w}(1-\bar z)^{2\bar h_w}}{z^{h_v+h_w} \bar z^{\bar h_v + \bar h_w}} \sum_{{\cal O}_s} C_{wvs}^2 \,{\cal V}_s(z,\bar z)\,.
\end{split}
\label{eq:crossing}
\end{equation}
The manipulations leading to this expression can be found in SM Sec.\ \ref{app:CFT_manips}.
The s-channel conformal blocks ${\cal V}_s(z,\bar z) = \langle {\cal B}_{WV{\cal O}_s}|{\cal B}_{WV{\cal O}_s} \rangle / \langle VV \rangle \langle WW \rangle$ are functions of conformal cross ratios, which are (for $ e^{t-|b|} \gg 1$):
\begin{equation}
z \approx 1 - 4\sin^2(\delta) e^{-t+b}\,,\quad
\bar z \approx 1 - 4\sin^2(\delta) e^{-t-b}\,,
\end{equation}
see SM Sec.\ \ref{app:CrossRatios} for details.
The conformal bootstrap program has established powerful tools to analyze the crossing equation \eqref{eq:crossing}. In particular, we use the results of Virasoro mean-field theory \cite{Ponsot:1999uf,Ponsot:2000mt,Jackson:2014nla,Chang:2016ftb,Caron-Huot:2017vep,Collier:2018exn}, which provides a precise account of the mean-field spectrum  ${\cal O}_s$ and the couplings $C_{wvs}^2$; see SM Sec.\ \ref{app:CFT_manips} for a summary. The relevant spectrum of ${\cal O}_s$ consists of two parts: 
\begin{enumerate}
    \item[$(i)$] A discrete spectrum of ``light'' double-twist operators ${\cal O}_s^{(m,\bar m)}$ with integer-spaced conformal weights
\begin{equation}
\label{eq:hmspec}
 (h_m,\, \bar h_{\bar m}) = (h_v + h_w + m,\, \bar h_v + \bar h_w + \bar m)\,,
\end{equation}
where $m = 1,2,\ldots,m_*$, and known OPE coefficients (similarly for $\bar m$). This spectrum receives anomalous corrections at ${\cal O}\big(\tfrac{1}{c}\big)$. The double-twist spectrum ends at the maximum values $(m_*,\bar m_*)$, where
\begin{equation}
\label{eq:hmax}
h_{m_*} = \bar{h}_{\bar m_*} = \frac{c-1}{24} \,.
\end{equation}
\item[$(ii)$] A continuous spectrum of ``heavy'' mean-field operators with conformal weights $(h_s,\bar{h}_s) > (h_{m_*},\bar{h}_{\bar m_*})$. Their average density of states and OPE coefficients are universally determined by the Virasoro fusion kernel. Heavy states are interpreted as dual to black holes with mass and spin \cite{Brown:1986nw,Keller:2014xba}
\begin{equation}
\label{eq:MJdef}
 M = h_s + \bar h_s - \frac{c}{12} \,,\qquad J = \bar h_s - h_s\,.
\end{equation} 
\end{enumerate}

Equation \eqref{eq:crossing} gives the probability distribution for the decomposition of the state \eqref{eq:OPEblocks} over Virasoro irreducible representations. Like in the one-shock case, boosting the operators probes higher descendant states within a given irrep. A novel feature of the two-shock setup is the fact that the state can now transition into different irreps. The latter effect is captured by the non trivial dependence of \eqref{eq:crossing} on Rindler time $t$. 

The central observation of this work is as follows: for small Rindler time separation $t\equiv-t_v-t_w$ (low boost), the s-channel decomposition is well described by the exchange of a localized superposition of Virasoro double-twist operators. The expected value of this {\it double-twist wave packet} grows exponentially in time. Black hole formation in the bulk corresponds to a breakdown of this approximation and the onset of heavy operator exchange.

In the next section, we justify this hypothesis by studying the light exchanges in detail. We show that the light states cease to dominate after a {\it black hole formation timescale}, i.e., when $t- |b|$ becomes of the order of twice the scrambling time. For previous discussions of this timescale, see \cite{Haehl:2017pak,Anous:2020vtw} in the context of scrambling and \cite{Haehl:2021prg,Haehl:2021tft} for shock wave collisions.

\section{Cross channel analysis: scrambling in-time-order}
\label{sec:analysis}

\subsection{Cross channel exchange: global limit}
\label{sec:analysis_global}

In this subsection we discuss precisely which operators ${\cal O}_s$ dominate in the cross-channel decomposition of ${\cal F}_\text{TOC}$ as a function of the kinematic parameters $(t,b,h_v,h_w,\delta)$. We focus on the early-time regime ($1 \ll e^{t-|b|} \ll c^2$), where only the discrete double-twist light states contribute to the OPE. In this regime, we can safely take the approximation $c\rightarrow \infty$ (referred to as ``global limit''), effectively extending the cutoff on the double-twist spectrum, $m_*,\bar{m}_* \rightarrow\infty$.

\begin{figure}
\includegraphics[width=.98\columnwidth]{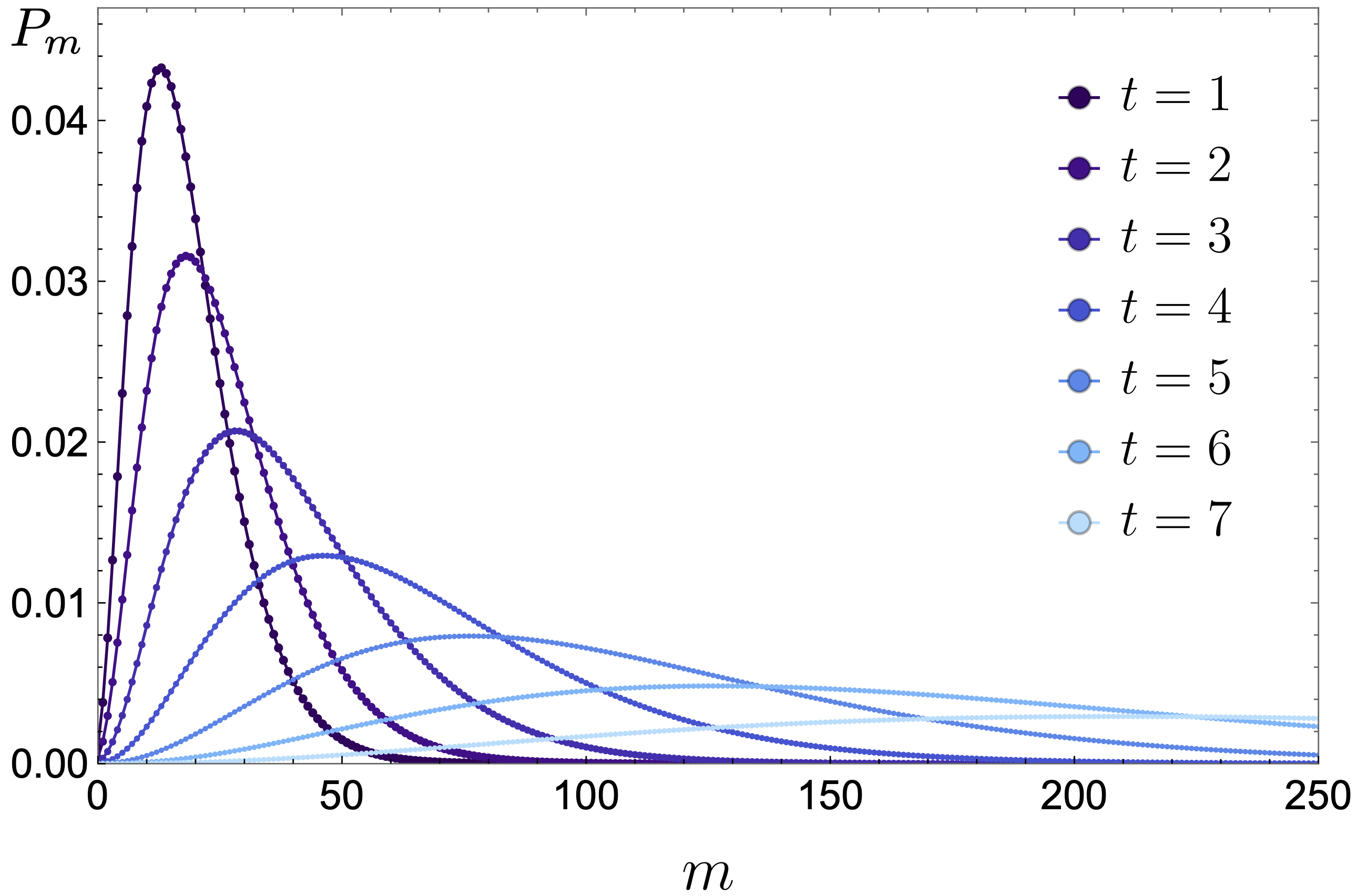}
\caption{The (discrete) probability distribution $P_m(h_v,h_w,z)$ on the space of possible s-channel exchange dimensions. The distribution is peaked and moves to higher weights exponentially with time $t$. We set $h_v=h_w=1,\, \delta=0.1,\,b=0$.}
\label{fig:wavepackets}
\end{figure}

The crossing equation \eqref{eq:crossing} can be viewed as the normalization condition for a probability distribution. In the global limit, this distribution is discrete:
\begin{equation}
\label{eq:normalization}
   1 = \left( \sum_{m \geq 0} P_m(h_v,h_w;z) \right) \left( \sum_{\bar m \geq 0} P_{\bar m}(\bar h_v,\bar h_w;\bar z) \right)\,,
\end{equation}
where $P_m$ is a probability distribution on the space of s-channel conformal blocks. That is, $P_m$ is the percentage contribution of ${\cal O}_s^{(m,\bar m)}$ defined by \eqref{eq:hmspec} (and its descendants) to the correlator ${\cal F}_\text{TOC} \approx 1$. 
In the global limit, we approximate the Virasoro double-twist conformal blocks by global conformal blocks, which take a simple analytic form (as reviewed in SM Sec.\ \ref{app:CFT_manips}):
\begin{equation}
\begin{split}
 P_m(h_v,h_w;z)&=(1-z)^{2h_w}\frac{z^m}{m!}\frac{(2h_v)_m (2h_w)_m}{(2h_m-m-1)_m} \\
 &\;\;\; \times \hypergeometricF{2}{1}{2h_w+m,2h_w+m}{2h_m}{\,z\,}\,.
\end{split}
\label{eq:PmDef}
\end{equation}  
Here, $P_m$ is the product of universal double-twist OPE coefficients $C_{wvs_m}$ and the global conformal blocks, which are functions of the cross ratio.

The identity \eqref{eq:normalization} and \eqref{eq:PmDef} is exact for any choice of kinematic parameters (proven in SM Sec.\ \ref{app:analytic}). However, the relevant range of $(h_m,\bar{h}_{\bar m})$ contributing support to the two sums is strongly dependent on the kinematics. Fig.\  \ref{fig:wavepackets} shows the probability distribution $P_m$ for fixed $(h_v,h_w,\delta)$ as a function of Rindler time difference $t=-t_v-t_w$. The distribution takes the form of a peaked wave packet; the peak moves toward larger values of $m$ exponentially in $t$, and it broadens at the same time. Let us now quantify this behavior.

The mean value of the exchanged operator dimension is $\mathbb{E}[ \Delta_s ] = \Delta_v + \Delta_w + \mathbb{E}[m]+\mathbb{E}[\bar m]$. This can be computed exactly (see SM Sec.\ \ref{app:analytic}). In the regime of interest ($ e^{t-|b|} \gg 1$), we find:
\begin{equation}
\label{eq:EmEmbar}
\begin{split}
\mathbb{E}[m] \equiv \sum_{m\geq 0} m  P_m &\approx \frac{1}{\sin(\delta)}\frac{\Gamma(2h_v+\tfrac{1}{2})\Gamma(2h_w+\tfrac{1}{2})}{2\Gamma(2h_v)\Gamma(2h_w)} \, e^{\frac{t-b}{2}} ,\\
\mathbb{E}[\bar m] \equiv \sum_{\bar m \geq 0}\bar m  P_{\bar m} &\approx \frac{1}{\sin(\delta)}\frac{\Gamma(2\bar h_v+\tfrac{1}{2})\Gamma(2\bar h_w+\tfrac{1}{2})}{2\Gamma(2\bar h_v)\Gamma(2\bar h_w)} \, e^{\frac{t+b}{2}}.
\end{split}
\end{equation}
The symbol $\mathbb{E}[\,\cdot\,]$ indicates a (statistical) expectation value with respect to the probability distribution $P_m$ on the space of quasiprimary exchange dimensions. Equivalently, one can formally define a ``size'' operator $\hat{m}$ which takes the value $m$ on any state in the representation $h_m$. We could then write the expectation value as $\mathbb{E}[m] = \langle \Psi_{W_LV_R} | \hat{m} | \Psi_{W_LV_R}\rangle/\langle V V\rangle \langle W W\rangle$.

\begin{figure}
\includegraphics[width=\columnwidth]{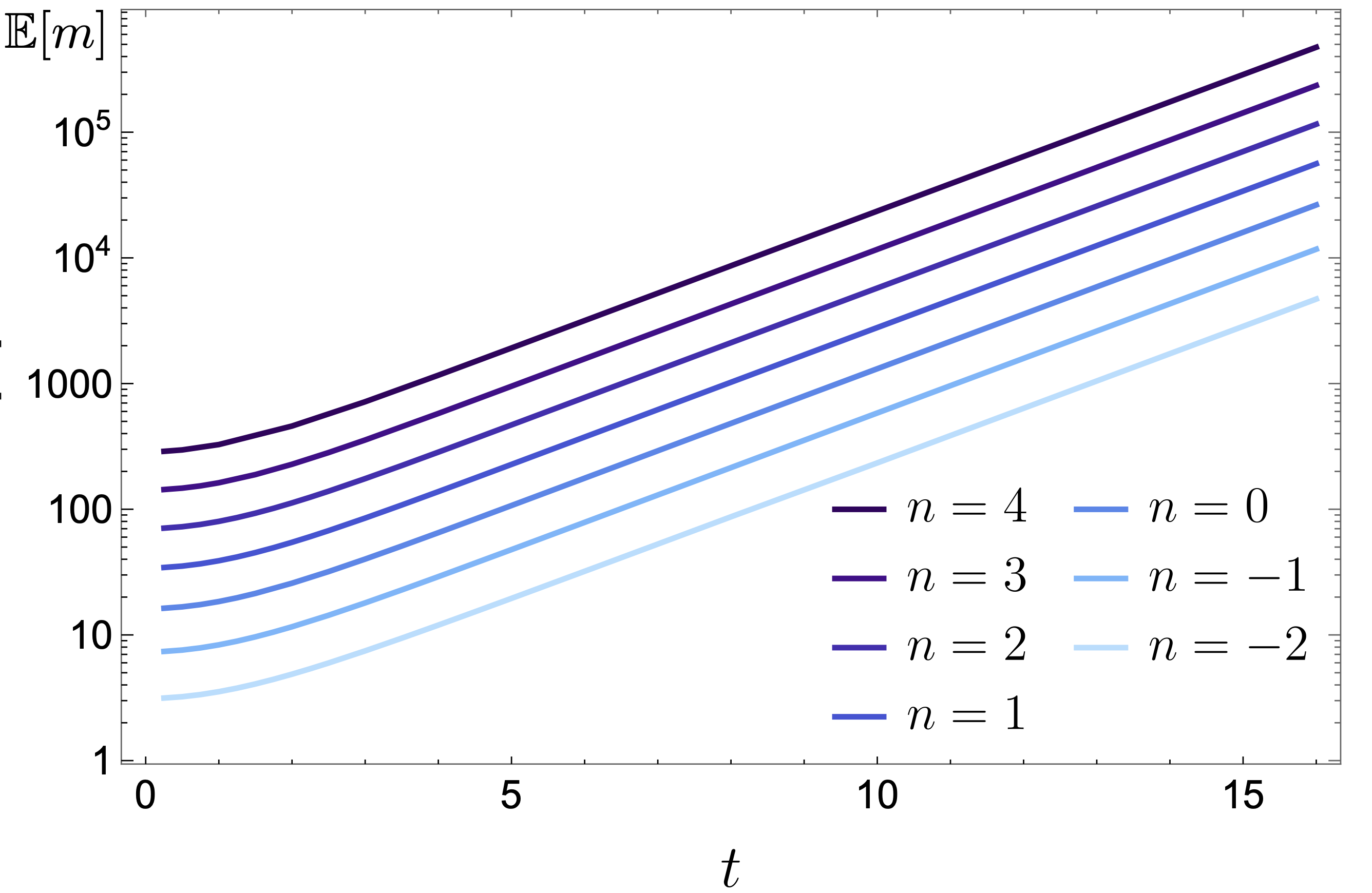}
\caption{The mean value of exchanged double-twist dimensions as a function of time. Different lines correspond to different external operator weights $h_v = h_w = 2^n$; $\delta=0.1$, $b=0$.}
\label{fig:means}
\end{figure}

In Fig.\  \ref{fig:means} we show the numerical evaluation of $\mathbb{E}[m]$: we observe the onset of exponential time dependence with a growth exponent $\frac{1}{2}$,  which is independent of the kinematic parameters and consistent with \eqref{eq:EmEmbar}.
The exponential growth of the mean exchanged operator dimension is a manifestation of the operator growth associated with the two-shock-wave state.

To further corroborate the operator growth picture, we note that the exponential increase of the mean exchanged dimension with time is accompanied by an exponential spreading of the probability distribution's width (see Fig.\  \ref{fig:wavepackets}). The two effects occur at an equal rate: the second moments of the distributions are (for $ e^{t-|b|} \gg 1$)
\begin{equation}
\sqrt{\mathbb{E}[m^2]} \approx \frac{\sqrt{h_v h_w}}{\sin(\delta)} \, e^{\frac{t-b}{2}} \,,\quad\;\;
\sqrt{\mathbb{E}[\bar{m}^2]} \approx \frac{\sqrt{\bar h_v \bar h_w}}{\sin(\delta)} \, e^{\frac{t+b}{2}} \,.
\end{equation}
These are of the same order as $\mathbb{E}[m]$ and $\mathbb{E}[\bar m]$. 
The tails of the distributions $P_m$ and $P_{\bar m}$, therefore, do not grow disproportionately, and we can sensibly identify a mean-localized wave packet even for large $t$. Higher moments are analyzed in the SM and behave similarly.

\subsection{Black hole formation}
\label{sec:BHformation}

Having seen how the light state s-channel support spreads exponentially in time, we now relax the assumption of infinite central charge and instead take it to be finite but large, $c\gg 1$. This restricts the validity of the global limit taken above.

For values of $t$ small compared to the scrambling time, the approximations of the previous subsection still hold. For simplicity, let us consider scalar operators with $h_v=\bar h_v \equiv \frac{1}{2}\Delta_v$ and $h_w=\bar h_w \equiv \frac{1}{2}\Delta_w$, and $1 \ll \Delta_{v,w} \ll c$. The Gamma functions in \eqref{eq:EmEmbar} can then be approximated using Stirling's formula. At early times, we propose that the bulk collision produces a conical defect geometry whose expected mass and spin are given by evaluating \eqref{eq:MJdef} on the expected values \eqref{eq:EmEmbar}:\footnote{We choose the ground state energy of global AdS$_3$ as $-\frac{c}{12}$. Below-threshold geometries thus have negative mass.
}
\begin{equation}
    \begin{split}
    M+\frac{c}{12} \equiv \mathbb{E}[\Delta_s] &\approx \frac{\sqrt{\Delta_v \Delta_w}}{\sin(\delta)} \, \cosh\left(\frac{b}{2}\right)\, e^{\frac{t}{2}}
    \,,\\
    J \equiv \,\mathbb{E}[\ell_s]\, &\approx \frac{\sqrt{\Delta_v \Delta_w}}{\sin(\delta)}  \, \sinh\left(\frac{b}{2}\right)\, e^{\frac{t}{2}}\,.
    \end{split}
\end{equation}
The threshold for black hole formation corresponds to an extremal geometry with mass $M=|J|$, cf.\ \cite{Banados:1992wn,Keller:2014xba}. Using the Brown-Henneaux relation $\frac{c}{12}=\frac{1}{8G_N}$, this can be written as \eqref{eq:gott2}. Once the threshold is reached, the black hole states dominate the s-channel exchange. The associated black hole states have mass $M>|J|$.

Let us understand this from the point of view of the mean-field theory spectrum for ${\cal O}_s$.
At finite $c$, the spectrum of discrete ``light'' double-twist operators ${\cal O}_s^{(m,\bar m)}$ ends sharply at the threshold $(m,\bar m)=(m_*(c),\bar{m}_*(c))$, corresponding to an extremal BTZ black hole, cf.\ \eqref{eq:hmax}. The threshold is reached when the mean of the wave packet of exchanged operators $(\mathbb{E}(m),\mathbb{E}(\bar m))$ attains values comparable to $\frac{c}{24}$, signaling breakdown of the global limit: the s-channel mean-field support transitions from discrete light states to a continuum of heavy states. These heavy states are precisely those describing the spectrum of BTZ black hole microstates. The onset of the breakdown of light state dominance amounts to the conditions:
\begin{equation}
\label{eq:Ethreshold}
    \text{BTZ threshold:} \qquad \text{min} \big\{ \mathbb{E}[m_*(c)] ,\, \mathbb{E}[\bar{m}_*(c)] \big\} \approx \frac{c}{24} \,. 
\end{equation}
In order to overcome the BTZ black hole threshold, it is important for both mean values to reach $\frac{c}{24}$, see \cite{Keller:2014xba}.\footnote{Our analysis here assumes that neither $\Delta_{v,w}$ nor $b$ scale with $c$.} Comparing with the functional dependence of $\mathbb{E}[m]$ and $\mathbb{E}[\bar m]$ in \eqref{eq:EmEmbar}, we can translate \eqref{eq:Ethreshold} (or, equivalently, the condition $M=|J|$) into a threshold timescale for black hole formation:
\begin{equation}
    t_\text{BH} -|b|\sim  2 \times \log \left( \frac{\sin(\delta)}{\sqrt{\Delta_v \Delta_w}} \; \frac{c}{12} \right) \,.
\end{equation}
We recognize this as twice the scrambling time, and as equivalent to the condition \eqref{eq:gott2} for black hole formation in gravity.

We comment briefly on the approximations made.
The threshold derived is not sharp, and 
some transient behavior takes place around the threshold time. When both $\mathbb{E}[m] \sim {\cal O}(\frac{c}{24}) \sim \mathbb{E}[\bar m]$, the spectrum $(h_m,\bar{h}_{\bar m})$ of light double-twist operators receives large anomalous corrections, and the Virasoro blocks are no longer well approximated by global blocks \cite{Collier:2018exn}. This is analogous to the physics of the OTOC around the scrambling time, which is also subject to transient behavior, leading to a breakdown of the large-$N$ approximation and of a simple exponential profile in time \cite{Roberts:2014ifa}.


\subsection{Distribution of descendants}
\label{sec:bdygravitons}

So far, we considered the average primary dimensions $(m,\bar m)$ of the s-channel wave packet as the parameter determining the exchanged state. This quantity is invariant, as it labels an irreducible orthogonal representation of the Virasoro algebra. Nevertheless, it is also interesting to consider the time dependent population of global conformal descendants within each such primary family.

In the global limit, for any given primary operator labeled by $(h_m,\bar{h}_{\bar m})$, the associated conformal block is built out of an infinite tower of global descendant operators with weights $(h_m+n,\bar{h}_{\bar m}+\bar n)$ for integers $n,\bar n \geq 0$.\footnote{Note that irreps of the global conformal group are labeled by the eigenvalue of the Casimir operator: ${\cal C} = L_0^2 - \frac{1}{2}(L_1L_{-1} +  L_{-1}L_1) = h_m(h_m-1)$.} This decomposition can be viewed as a joint probability distribution $P_{m,n}(h_v,h_w;z)$, which breaks up the primary distribution $P_{m}(h_v,h_w;z)$ into descendant levels (see SM Sec.\ \ref{app:analytic_descendants}). Note that this decomposition depends on the choice of a quantization scheme: we need to specify around which point the conformal blocks are expanded, i.e., a reference point for the action of $L_0+\bar L_0$. We choose the canonical expansion into descendants with respect to $z=\bar z=0$ and indicate this with a superscript:
\begin{equation}
    P_{m}(h_v,h_w;z) = \sum_{n\geq 0} P^{(V)}_{m,n}(h_v,h_w;z)\,,
\end{equation}
and similarly for $P_{\bar{m}}$.
The choice of expansion point and map to a canonical configuration in the CFT corresponds to a choice of frame in gravity. The distribution of global descendants depends on these choices.

\begin{figure}
\includegraphics[width=\columnwidth]{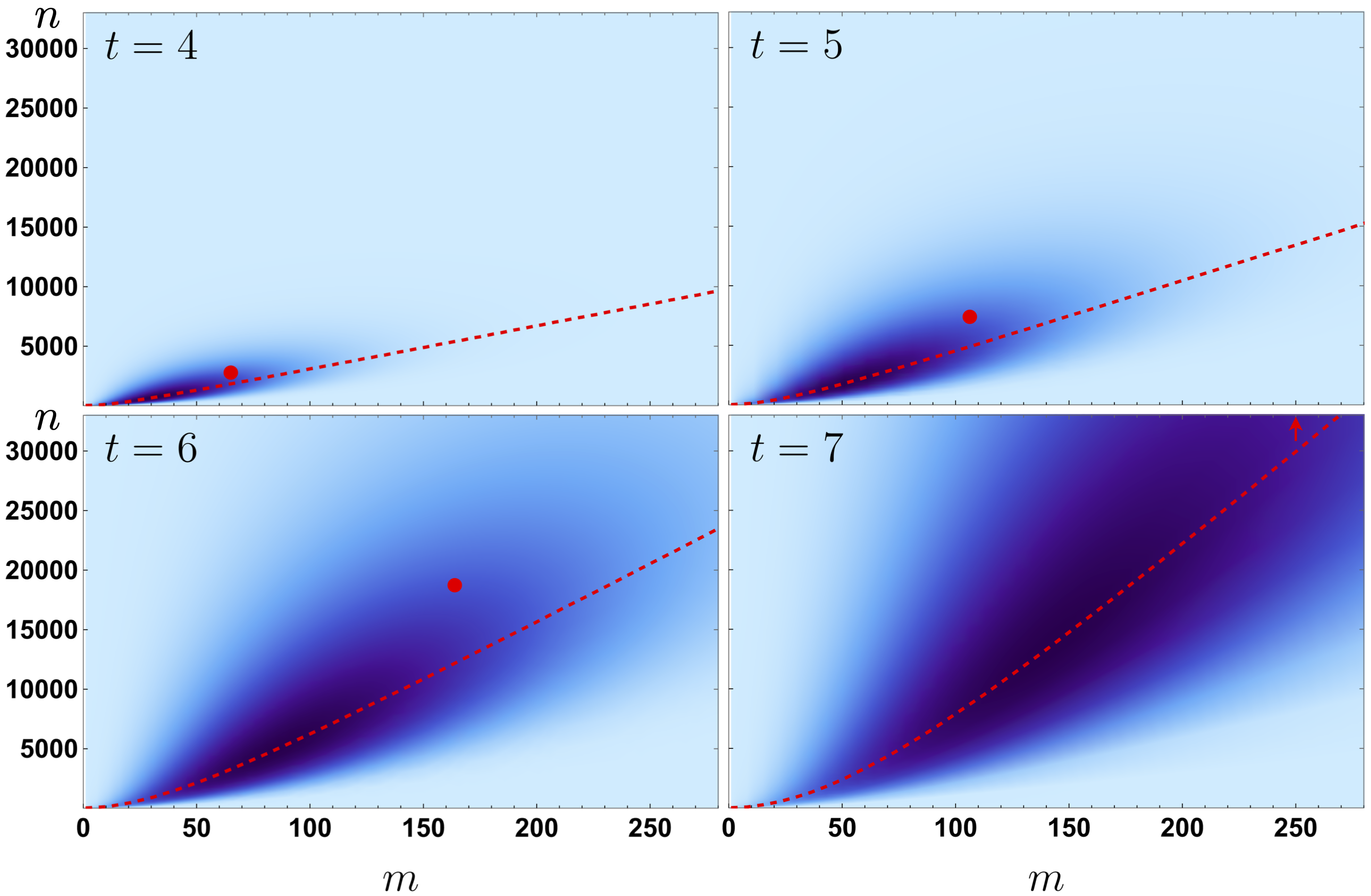}
\caption{The joint distribution $P^{(V)}_{m,n}(h_v,h_w,z)$ for $h_v=h_w=1$ at different times $t$. Darker colors indicate more support. The dashed lines indicate peak values in the vertical direction for fixed double-twist primary $h_m$. The dots mark the point $(\mathbb{E}[m],\mathbb{E}^{(V)}[n])$, which determines the global energy $\mathbb{E}^{(V)}[L_0]$.}
\label{fig:globalenergy}
\end{figure}

In the large-$c$ approximation of Virasoro conformal blocks by global  blocks, the structure of the global descendants (i.e., the distribution $P^{(V)}_{m,n}$) is well known \cite{Dolan:2000ut}; it corresponds to the series expansion of the hypergeometric function \eqref{eq:PmDef} in powers of the cross ratio. The $n$-th descendant level is characterized by a quantum number $h_{m,n}=h_v+h_w+m+n$ under the global time translation generator $L_0$ (similarly for $\bar L_0$). We can, thus, compute the expectation value of the global energy of the exchanged states analytically. We find (for $c\gg e^{t-|b|} \gg 1$):
\begin{equation}
\label{eq:Eglobal}
\begin{split}
 \mathbb{E}^{(V)}[L_0] = \sum_{m,n}(h_m+n)P^{(V)}_{m,n} &\approx h_v + h_w +  \frac{h_w}{2\sin^2(\delta)}  e^{t-b}  ,\\
 \mathbb{E}^{(V)}[\bar{L}_0]= \sum_{\bar m,\bar n}({\bar h}_{\bar m}+\bar n)P^{(V)}_{\bar m,\bar n} &\approx \bar h_v + \bar h_w +  \frac{\bar h_w}{2\sin^2(\delta)}  e^{t+b}.
 \end{split}
\end{equation} 
Notably, this growth proceeds at twice the rate of the growth of $\mathbb{E}[m]$ or $\mathbb{E}[\bar m]$. We derive this in SM Sec.\ \ref{app:analytic_descendants} and illustrate it in Fig.\  \ref{fig:globalenergy},
which depicts the growth of the exchanged operator in the two-shock-wave state over the space of possible s-channel primaries (labeled by $m$) and global descendants (labeled by $n$). A similar notion of global energy increase due to operator growth into descendant levels under Rindler time evolution was previously discussed for a single shock in \cite{Caputa:2021ori}.

The global energy of the exchanged state, \eqref{eq:Eglobal}, becomes of the order of $\frac{c}{12}$ after one scrambling time $t_*$. In gravity, this marks the point where gravitational backreaction can no longer be ignored: the spacetime region in the future light-cone of the collision point begins to {\it shrink} due to strong gravity effects \cite{Haehl:2022frr}.\footnote{This effect can also be understood using quantum circuit models \cite{Haehl:2022frr}, which implement the interplay between the spreading of each precursor operator's effect inside a causal light-cone and the ballistic spread of its scrambling dynamics inside a ``butterfly cone'' \cite{Roberts:2014isa,Mezei:2019dfv}.
}

\section{Conclusion}
\label{sec:conclusion}

In this Letter we proposed a precise CFT manifestation of black hole formation in AdS$_3$ via colliding shock waves: we identified a wave packet of mean-field operators exchanged in the cross-channel operator product expansion of two highly boosted precursor operators. The expected value of the distribution of exchanged operators crosses the BTZ black hole threshold at a timescale equal to twice the scrambling time, i.e., when both precursors are sufficiently boosted to generate shock waves. By phrasing this in terms of operator growth within the space of conformal blocks, we discovered scrambling dynamics -- normally associated with out-of-time-order correlators -- using an in-time-order four-point function diagnostic.

In the future, we plan to report on a detailed analysis of the operator product expansion for timescales exceeding the black hole formation threshold. In this regime the exchange is dominated by heavy states associated with black holes \cite{Das:2017cnv,Collier:2018exn}. We expect a qualitatively different distribution of energy among the descendant operators: excitations of Virasoro (as opposed to global) modes are no longer suppressed by powers of $\frac{1}{c}$. These excitations are interpreted as genuine gravitational dressing \cite{Keller:2014xba} that cannot be absorbed into global isometry transformations. It is interesting to study the fraction of energy carried by these boundary gravitons, and whether this effect translates into a localization of the primary distribution, sharpening the black hole formation transition.

Furthermore, a detailed CFT characterization of the causal structure of the bulk geometry and horizon physics is desirable. The simplicity of geometric horizon formation in gravity suggests that there might be more primitive CFT probes, related to the bulk causal structure, which would detect the same process.

We hope that these insights will pave the way toward a precise definition of operator growth and operator complexity in quantum field theory. Via holography, a detailed microscopic mechanism describing black hole formation may offer key insights into strongly backreacting gravitational dynamics and singularities.

\vspace{1cm}

\begin{acknowledgments}
The authors are grateful to V.\ Balasubramanian, A.\ Belin, C.-M.\ Chang, C.\ Chowdhury, S.\ Collier, B.\ Czech, T.\ Hartman, V.\ Pellizzani, E.\ Perlmutter, E.\ Rabinovici, M.\ Rangamani, A.\ Rolph, K.\ Skenderis, J.\ Sonner, P.\ Tadic and B. Withers for enlightening discussions. FMH and ASG are supported by UK Research and Innovation (UKRI) under the UK government’s Horizon Europe Funding Guarantee EP/X030334/1. YZ is supported by DOE High Energy DE-SC0012567, and  DOE QIS  DE-SC0025937. This research was supported in part by grant NSF PHY-2309135 to the Kavli Institute for Theoretical Physics (KITP).
\end{acknowledgments}

\newpage

\appendix

\onecolumngrid

\section*{SUPPLEMENTAL MATERIAL}

\section{DETAILS ON THE KINEMATIC SETUP}\label{app:CrossRatios}

In this appendix we give details about the kinematic setup for computing the time-ordered four-point function discussed in the main text. The exact expression of the correlator is:
\begin{equation}
    \label{eq:Correlator_withzzb}
    \mathcal{F}_{\rm TOC}(z,\bar{z}) =\frac{\langle V(z_1,\bar z_1)W(z_2,\bar z_2)W(z_3,\bar z_3)V(z_4,\bar z_4) \rangle}{\langle V(z_1,\bar z_1)V(z_4,\bar z_4)\rangle\langle W(z_2,\bar z_2) W(z_3,\bar z_3)\rangle}~, \qquad (z,\bar z)=\left(\frac{z_{12} z_{34}}{z_{13}z_{24}} ,\, \frac{\bar z_{12} \bar z_{34}}{\bar z_{13}\bar z_{24}}\right) \,,
\end{equation}
where the expectation values are taken in the global vacuum and the normalized correlator depends on the cross ratios $(z,\bar z)$. The insertion points on the plane $(z_i,\,\bar{z}_i)$ are related to the Rindler parametrization via 
\begin{equation}
    \label{eq:Rindler_to_place}
    z_i=e^{t_i+x_i}~,\qquad \bar{z}_i=e^{-t_i+x_i}~.
\end{equation}
In our two-shock-wave setup, operators in the norm of the state \eqref{eq:Two_shock_state} are placed as follows:
\begin{align}
    & ~V:~~\;(t_1,x_1)=(t_v-i\delta,x_v)\,,\qquad\qquad W:~~(t_2,x_2)=(-t_w-i\pi +i\delta,x_w) \,, \label{eq:Operator_insertions_ket}\\
    & W:~~~(t_3,x_3)=(-t_w+i\pi-i\delta,x_w)\,,\quad V:~~(t_4,x_4)=(t_v+i\delta,x_v)~.\label{eq:Operator_insertions_bra}
\end{align}
Defining the precursor time $t:=-t_v-t_w>0$ and the impact parameter $b:=x_v-x_w\in\mathbb{R}$,\footnote{Our definition of impact parameter aligns with that in \cite{Holst:1999tc,Haehl:2023lfo}, where it is taken to be the distance between the colliding shocks, rather than the distance between each individual shock and the center of mass.} the cross ratios are
\begin{equation}
    \label{eq:Cross_ratios_exact}
    z = 1-\frac{2 \sin ^2(\delta )}{1+\cosh (t-b)} \approx 1 - 4 \sin^2(\delta) \, e^{-(t-b)} 
    ~,\qquad \bar{z} = 1-\frac{2 \sin ^2(\delta )}{1+\cosh (t+b)} \approx 1 - 4 \sin^2(\delta) \, e^{-(t+b)} ~.
\end{equation}
where the approximations are valid for $e^{t-|b|} \gg  1$.
We note that both $z$ and $\bar{z}$ are real and always within the region of convergence of the OPE. This is a manifestation of the fact that $\mathcal{F}_{\rm TOC}(z,\bar{z})$ is a time-ordered correlation function.

\section{CROSSING EQUATION AND MEAN FIELD SPECTRUM}\label{app:CFT_manips}

This appendix summarizes the manipulations used to obtain the decomposition of the norm of the two-shock state $|\Psi_{WV}\rangle$ in terms of orthogonal contributions coming from different conformal families weighted by the probabilities $P_m(h_v,h_w;z)$. The main tool we shall exploit is the Virasoro fusion kernel, see \cite{Ponsot:1999uf,Ponsot:2000mt} and the more recent \cite{Collier:2018exn,Kusuki:2024gtq} (which we follow closely). The two physical assumptions in the computation will be $(i)$ Virasoro identity block dominance and $(ii)$  approximation of Virasoro blocks by global conformal blocks at early times (``global limit'').

Crossing symmetry implies that the normalized time-ordered four-point function \eqref{eq:Correlator_withzzb} can be expanded into Virasoro blocks in two channels (s- and t-channel):
\begin{align}
    \mathcal{F}_{\rm TOC}(z,\bar{z}) &= \frac{(1-z)^{2h_w}}{z^{h_v+h_w}} \frac{(1-\bar{z})^{2\bar{h}_w}}{\bar{z}^{\bar{h}_v+\bar{h}_w}} \sum_{h_s,\bar{h}_s}C_{wvs}\,C^{s}_{~wv} \,\mathcal{V}^{h_w,h_v}_{h_w,h_v}(h_s;z) \,\mathcal{V}^{\bar{h}_w,\bar{h}_v}_{\bar{h}_w,\bar{h}_v}(\bar{h}_s;\bar{z}) \label{eq:Conformal_block_Expansion_Conventions_s} \\
    &= \sum_{h_t,\bar{h}_t}C_{vvt}\,C^{t}_{~ww}\, \mathcal{V}^{h_w,h_w}_{h_v,h_v}(h_t;1-z) \,\mathcal{V}^{\bar{h}_w,\bar{h}_w}_{\bar{h}_v,\bar{h}_v}(\bar{h}_t;1-\bar{z})~, \label{eq:Conformal_block_Expansion_Conventions_t}
\end{align}
where the indices in the OPE coefficients $C_{ijk}$ are raised and lowered with the Zamolodchikov metric \cite{Zamolodchikov:426555}, which can be chosen to be flat amongst Virasoro primaries. 
The blocks are normalized such that $\mathcal{V}^{h_2,h_1}_{h_3,h_4}(h,z)\sim z^{h}$ near $z\sim 0$, and the two-point function normalization in \eqref{eq:Correlator_withzzb} has been taken into account in the overall prefactor. As explained in the main text, the s-channel blocks admit the interpretation of orthogonal contributions to the norm of the two-shock-wave state $|\Psi_{WV}\rangle$. The crossing equation allows us to constrain these by considering the complementary t-channel expansion.
In particular, in holographic CFTs with a large central charge and a gapped spectrum, it is natural to assume that the t-channel decomposition is dominated by the identity operator \cite{Hartman:2013mia}:
\begin{equation}
    \label{eq:Identity_dominance}
    \mathcal{F}_{\rm TOC}(z,\bar{z})\approx \mathcal{V}^{h_w,h_w}_{h_v,h_v}(h_t=0;1-z) \,\mathcal{V}^{\bar{h}_w,\bar{h}_w}_{\bar{h}_v,\bar{h}_v}(\bar h_t = 0;1-\bar{z})\,,
\end{equation}
plus terms suppressed by the central charge. 

The blocks in the s- and t-channels can be related by the so-called Virasoro fusion kernel $\mathbb{S}$, which is fully determined by the Virasoro algebra \cite{Ponsot:1999uf,Ponsot:2000mt}. For identity exchange in the t-channel ($h_t=0$) and arbitrary central charge $c$, the relation takes the following form:
\begin{equation}
\label{eq:Fusion_kernel_generic_but_schematic}
    \frac{z^{h_v+h_w}}{(1-z)^{2h_w}}\,\mathcal{V}^{h_w,h_w}_{h_v,h_v}(0;1-z) = \sum_{m=0}^{m_*} R_m ~\mathcal{V}^{h_w,h_v}_{h_w,h_v}(h_m,z)+ \int_{\frac{c-1}{24}}^{\infty}d\mu(h_s)~\mathbb{S}\left[\begin{matrix}
        h_w &h_v\\
        h_w & h_v
    \end{matrix}\right]_{h_s,h_t=0} ~\mathcal{V}^{h_w,h_v}_{h_w,h_v}(h_s;z)~.
\end{equation}
The sum should be viewed as accounting for the ``light'' spectrum of double-twist mean-field operators; the integral covers the spectrum of ``heavy'' operators above the black hole threshold.
We refer the reader to \cite{Collier:2018exn} for details on the form of the kernel and the measure over the heavy states, which takes a simple form using a Liouville-like parametrization. When the external dimensions $h_v$ and $h_w$ are below the black hole threshold (as we assume throughout), the light spectrum is given by $h_m=h_v+h_w+m-\delta m$, where $m=0,\dots,m_{*}$, and $\delta m$ is a positive, finite-central-charge correction that vanishes in the global limit $c\rightarrow\infty$. The upper bound $m_*$ is such that $h_m\leq \frac{c-1}{24}$. The light states are weighted by coefficients $R_m$ related to the singularity structure of the kernel $\mathbb{S}$ \cite{Collier:2018exn}. Identity block dominance in the t-channel \eqref{eq:Identity_dominance} then implies:
\begin{align}
    \mathcal{F}_{\rm TOC}(z,\bar{z}) &\overset{\eqref{eq:Identity_dominance}}{\approx}\mathcal{V}^{h_w,h_w}_{h_v,h_v}(0;1-z) \,\mathcal{V}^{\bar{h}_w,\bar{h}_w}_{\bar{h}_v,\bar{h}_v}(0;1-\bar{z}) \nonumber\\
    &\overset{\text{\eqref{eq:Fusion_kernel_generic_but_schematic}}}{=} \frac{(1-z)^{2h_w}}{z^{h_v+h_w}}\left(\sum_{m=0}^{m_*} R_m ~\mathcal{V}^{h_w,h_v}_{h_w,h_v}(h_m;z)+ \int_{\frac{c-1}{24}}^{\infty}d\mu(h_s)~\mathbb{S}\left[\begin{matrix}
        h_w &h_v\\
        h_w & h_v
    \end{matrix}\right]_{h_s,h_t=0} ~\mathcal{V}^{h_w,h_v}_{h_w,h_v}(h_s;z)\right)\times\big[ \text{anti-holo.} \big],
    \label{eq:FTOC_identity_dominance_and VMFT_line2}
\end{align}
where the anti-holomorphic factor is structurally identical to the holomorphic one. We stress that the second equality is an exact identity, only based on symmetry. Comparing \eqref{eq:FTOC_identity_dominance_and VMFT_line2} with \eqref{eq:Conformal_block_Expansion_Conventions_s}, one may understand the coefficients $R_m$ and the kernel $\mathbb{S}[\,\cdot\,]_{h_s,0}$ as an effective, or mean-field, description of the s-channel OPE coefficients for light and heavy exchanged operators, respectively, in a theory dual to semiclassical gravity.\footnote{See also \cite{Collier:2019weq,Belin:2020hea} for related ideas in the context of holography.}

Given that in our setup the external operators sourcing the bulk shock waves are light, the dominant exchanged dimensions in the s-channel will also be light at early times. It is thus justified to study early-time dynamics by considering the global limit of \eqref{eq:FTOC_identity_dominance_and VMFT_line2}. In particular, in this limit we take $c\to\infty$ independent of both the external and the internal dimensions. In this limit, the integral contribution in \eqref{eq:Fusion_kernel_generic_but_schematic} disappears, provided that its integrand is a decaying function of $h_s$, while the mean-field OPE coefficients take the form \cite{Fitzpatrick:2011dm,Collier:2018exn}:
\begin{equation}
    \label{eq:Global_Limit_residues}
    \lim_{c\to\infty}R_m = \frac{(2h_v)_m (2h_w)_m}{m!\, (2h_v+2h_w-1+m)_m}~,
\end{equation}
where $(a)_n=\Gamma(a+n)/\Gamma(a)$.
Additionally, the Virasoro blocks become global blocks in this limit. In particular,
\begin{align}
    \lim_{c\to \infty}\mathcal{V}^{h_w h_v}_{h_w h_v}(h_m;z) &= z^{h_m}\hypergeometricF{2}{1}{2h_w+m,~2h_w+m}{2h_v+2h_w+2m}{z}~, \label{eq:Global_s_block} \\
    \lim_{c\to \infty}\mathcal{V}^{h_w h_w}_{h_v h_v}(0;1-z) &=1~,\qquad\text{for any }z\in[0,1)~, \label{eq:global_t_block_is_one}
\end{align}
where $h_m = h_v+h_w+m$ in the global limit.
Plugging \eqref{eq:Global_Limit_residues}-\eqref{eq:global_t_block_is_one} in \eqref{eq:FTOC_identity_dominance_and VMFT_line2}, we obtain:
\begin{equation}
    \label{eq:FTOC_decomposition_global}
    \mathcal{F}_{\rm TOC}(z,\bar{z})\overset{\substack{\text{id.\ dominance} \\ +\,\text{global limit}}}{\approx} 1 = \sum_{m=0}^{\infty} P_m(h_v,h_w;z) \sum_{\bar{m}=0}^{\infty}P_{\bar{m}}(\bar{h}_v,\bar{h}_w;\bar{z})~,
\end{equation}
where we have identified the probabilities weighting the orthogonal contributions to $\mathcal{F}_{\rm TOC}$ as:
\begin{equation}
    \label{eq:Probabilities_original_formulation}
    P_m(h_v,h_w;z):=(1-z)^{2h_w}\,\frac{z^m}{m!}\frac{(2h_v)_m (2h_w)_m}{(2h_v+2h_w+m-1)_m} \hypergeometricF{2}{1}{2h_w+m,~2h_w+m}{2h_v+2h_w+2m}{z}~.
\end{equation}
As we show in SM Sec.\ \ref{app:analytic}, the fact that the sums in \eqref{eq:FTOC_decomposition_global} are equal to $1$ can be independently proved using hypergeometric function manipulations.

\section{ANALYTICAL ANALYSIS OF SUM OVER DOUBLE-TWISTS}
\label{app:analytic}

This appendix provides the analytical expressions for the probabilities $P_m$ used in Sec.\  \ref{sec:analysis} and their moments.\footnote{The authors thank Vito Pellizzani for valuable suggestions on hypergeometric identities that facilitated this analysis.} We also use these results to analyze the descendant states that control the global energy discussed in Sec.\  \ref{sec:bdygravitons}.

\subsection{Distribution over conformal families}

We decompose the (holomorphic part of the) CFT Hilbert space into orthogonal subspaces labeled by primary operators. The probability $P_m(h_v,h_w;z)$ gives the contribution to the norm of the two-shock-wave state \eqref{eq:FTOCdef} due to the quasiprimary ${\cal O}_s^{(m)}$ with dimension $h_m=h_v+h_w+m$ and its descendants. It combines the relevant kinematic factors, double-trace OPE coefficients, and global conformal block. It will be convenient to rewrite \eqref{eq:Probabilities_original_formulation} as follows:
\begin{equation}
    \label{eq:Prob_compact}
    \begin{split}
    P_m(h_v,h_w;z)&=
    \text{probability of populating the conf.\ irrep. } h_m \text{ within } |\Psi_{WV}\rangle  \\
    &=\frac{1}{m!}\left(\frac{z}{1-z}\right)^m \frac{(2h_v)_m (2h_w)_m}{(2h_v+2h_w+m-1)_m} \hypergeometricF{2}{1}{2h_v+m,2h_w+m}{2h_v+2h_w+2m}{\frac{z}{z-1}}\,.
    \end{split}
\end{equation}
This distribution can be characterized via its moments $\mathbb{E}[m^k]:=\sum_{m\geq 0} m^k~P_m(h_v,h_w;z)$ for $k\geq 0$. To compute these moments, we write $m^k$ in terms of falling factorials $m^{\underline{p}}= \Gamma(m+1)/\Gamma(m+1-p)$ and Stirling numbers of the second kind:
\begin{equation}
    m^k = \sum_{p=0}^k {\cal S}(k,p)\,m^{\underline{p}} \,,\qquad \mathcal{S}(k,p) = \sum_{\ell=0}^p \frac{(-1)^{p-\ell} \ell^k}{(p-\ell)! \ell!}  \,.
\end{equation}
The desired $k$-th moment of the distribution $P_m$ is then given by the sum over expectation values of falling factorials:
\begin{equation}
    \label{eq:Moments_from_falling_factorials}
    \mathbb{E}[m^k]=\sum_{p=0}^k \mathcal{S}(k,p) \, \mathbb{E}[m^{\underline{p}}] 
    \,,\qquad 
    \mathbb{E}[m^{\underline{p}}] := \sum_{m=p}^\infty  m^{\underline{p}} \, P_m(h_v,h_w;z)\,.
\end{equation}
To compute these moments, it is useful to derive a simple expression for the sum
\begin{align}
    S_p (a,b,c;x)&= \sum_{m=p}^{\infty} \frac{(-x)^m}{(m-p)!} \frac{(a)_m~ (b)_m}{(c+m-1)_m} \hypergeometricF{2}{1}{a+m,b+m}{c+2m}{x} \,,\label{eq:Falling_fact_sum_def} 
\end{align}
which encodes the expectation values of the falling factorials appearing in \eqref{eq:Moments_from_falling_factorials}, by evaluation on the following special values:
\begin{align}
\label{eq:Expmp}
    \mathbb{E}[m^{\underline{p}}] = S_p \left(2h_v,2h_w,2h_v+2h_w;\frac{z}{z-1}\right)    \,.
\end{align}
In order to compute the sum \eqref{eq:Falling_fact_sum_def}, one may start by shifting the summation index $m\mapsto m+p$ so that the range of the new index starts at zero. Taylor-expanding the hypergeometric function in the summand and collecting powers of $x$, one reaches the following series:
\begin{equation}
\label{eq:Proof_HyperIdentity_step1}
    S_p(a,b,c;x) = (-x)^p\frac{ (a)_p  (b)_p (c-1)_p}{(c-1)_{2p}} \sum_{k=0}^{\infty} x^k \frac{(a+p)_k~(b+p)_k}{k! (c+2p)_k} \hypergeometricF{3}{2}{-k,~\frac{c+2p+1}{2},~c+p-1}{\frac{c+2p-1}{2},~c+2p+k}{1}~.
\end{equation}
The hypergeometric function above can be simplified by noting that the second upstairs argument differs by $1$ from the first downstairs argument, which allows to rewrite it as:\footnote{Cf. \href{https://functions.wolfram.com/HypergeometricFunctions/Hypergeometric3F2/03/02/02/0001/}{https://functions.wolfram.com/HypergeometricFunctions/Hypergeometric3F2/03/02/02/0001/}.}
\begin{equation}
\label{eq:Proof_HyperIdentity_step2}
    \hypergeometricF{3}{2}{-k,~\frac{c+2p+1}{2},~c+p-1}{\frac{c+2p-1}{2},~c+2p+k}{1} = \frac{p}{(c +2p-1)} \frac{ \Gamma (2 k+p) \Gamma (k+2p+c )}{ \Gamma (k+p+1) \Gamma (2 k+2p+c -1)}~.
\end{equation}
Expressing the Gamma functions in \eqref{eq:Proof_HyperIdentity_step2} in terms of Pochhammer symbols allows to directly identify the sum in \eqref{eq:Proof_HyperIdentity_step1} as the Taylor series of a hypergeometric function $_4F_3$: 
\begin{align}
    S_p (a,b,c;x)= (-x)^p \frac{ (a)_p  (b)_p (c-1)_p}{(c-1)_{2p}}  \hypergeometricF{4}{3}{\frac{p+1}{2},~\frac{p}{2},~a+p,~b+p}{p+1,~\frac{c-1}{2}+p,~\frac{c}{2}+p}{~x} \label{eq:Falling_fact_sum_result}~.
\end{align}

The result \eqref{eq:Falling_fact_sum_result} contains all information about the moments of the distribution $P_m$ via \eqref{eq:Moments_from_falling_factorials} and \eqref{eq:Expmp}. 
For example, by explicit evaluation for $k=0,1,2$, we find the first few moments:
\begin{align}
    & \mathbb{E}[1] = 1~, \label{eq:Zeroth_moment} \\
    &\mathbb{E}[m]= \frac{1}{2}\left(2h_v+2h_w-1\right)\left\{-1+\hypergeometricF{3}{2}{-\frac{1}{2},~2h_v,~2h_w}{h_v+h_w-\frac{1}{2},~h_v+h_w}{\frac{z}{z-1}}\right\}~,  \label{eq:First_moment}\\
    & \mathbb{E}[m^2] = 4h_v h_w\,\frac{z}{1-z} - (2h_v+2h_w-1)\,\mathbb{E}[m]~.  \label{eq:Second_moment}
\end{align}
We note that \eqref{eq:Zeroth_moment} states the normalization of the distribution $P_m(h_v,h_w;z)$; \eqref{eq:First_moment} gives the average conformal weight contributing to the two-shock-wave state; and the second moment \eqref{eq:Second_moment} characterizes the spread of the distribution.

\subsection{Distribution of global descendants}
\label{app:analytic_descendants}

As explained in the main text, besides spreading across conformal families of progressively higher primary dimension, in the global limit the two-shock state also spreads as a function of the precursor time $t$ over the descendants belonging to the global families that contribute to it. This is measured by the expectation value of the global energy operator $L_0+\bar{L}_0$.
In this section we provide the computation of the expectation value $\mathbb{E}[L_0] \equiv \langle\Psi_{WV}| L_0|\Psi_{WV}\rangle / \langle VV\rangle \langle WW \rangle$ in the two shock state (the computation of $\mathbb{E}[\bar{L}_0]$ is analogous).

In order to decompose the two-shock state as a linear combination over orthogonal projections over global families, a quantization scheme needs to be specified: we obtain the decomposition by applying an operator product expansion to the product $WV$ in the correlator
\begin{equation}
    \label{eq:Correlator_Canonical}
    \mathcal{F}_{\rm TOC}(z,\bar{z}) =\frac{\langle V(\infty,\infty)W(1,1)W(z,\bar z)V(0,0) \rangle}{\langle V(\infty,\infty)V(0,0)\rangle\langle W(1,1) W(z,\bar z)\rangle}~,
\end{equation}
which is related to \eqref{eq:Correlator_withzzb} by a global conformal transformation, and where the $s$-channel expansion is around $z\to 0$. This amounts to choosing a radial quantization scheme whose origin coincides with the location of the $V$ operator. Physically, this is equivalent to considering a kinematic setup equivalent to that presented in SM Sec.\ \ref{app:CrossRatios}, where instead of boosting both operators relative to the $t_R=0$ slice, we fix the $V$-insertion and only boost $W$ relative to it.\footnote{Indeed, note that the cross ratios \eqref{eq:Cross_ratios_exact} only depend on the combination $t=-(t_v+t_w)$ rather than separately on $t_v$ and $t_w$.} As derived in SM Sec.\ \ref{app:CFT_manips}, the global block corresponding to this OPE channel is given by \eqref{eq:Global_s_block} with quasiprimary dimensions $(h_m,\bar h_{\bar m})=(h_v+h_w+m,\bar h_v+\bar h_w+\bar m)$.
To extract the contribution of level-$n$ descendants of the state labeled by $h_m$, consider 
the global conformal symmetry generators satisfying $[L_0,L_{\pm 1}] = \mp L_{\pm 1}$ and $[L_1,L_{-1}]=2L_0$.
Since $|h_m\rangle$ is the quasiprimary state characterized by $L_0 |h_m\rangle = h_m |h_m\rangle$, a level-$n$ descendant  within its conformal family is given by
\begin{equation}
    |h_{m,n}\rangle = \frac{1}{\sqrt{N_{m,n}}} \, L_{-1}^n |h_m\rangle\,, \qquad  h_{m,n} = h_m + n\,,
\end{equation}
where the normalization evaluates to $N_{m,n}\equiv \langle h_m | L_1^n L_{-1}^n |h_m\rangle=n! \, (2h_m)_n$.
The conformal weight of the descendant is $L_0 |h_{m,n}\rangle = h_{m,n} |h_{m,n}\rangle$. Similar considerations apply to $|\bar{h}_{\bar m}\rangle$.

We can use these observations to define a projector onto the exchange of level $(n,\bar n)$ descendants of the quasiprimary labeled by $(h_m,\bar h_{\bar m})$:
\begin{equation}
\label{eq:PiDef}
  \Pi_{h_m,\bar h_{\bar m}}^{(n,\bar n)} =   \Pi_{h_m}^{(n)}   \otimes  \Pi_{\bar h_{\bar m}}^{(\bar n)}  = \left(  \frac{ L_{-1}^n | h_m \rangle \langle h_m | L_1^n }{\langle h_m | L_1^n L_{-1}^n |h_m\rangle}  \right) \otimes\left(  \frac{ \bar L_{-1}^n | \bar h_{\bar m} \rangle \langle \bar h_{\bar m} | \bar L_1^n }{\langle \bar h_{\bar m} | \bar L_1^n \bar L_{-1}^n |\bar  h_{\bar m}\rangle}  \right) \,.
\end{equation}
By inserting the projector \eqref{eq:PiDef} into ${\cal F}_\text{TOC}$, we clearly obtain a probability distribution over the space of global descendants of ${\cal O}_s^{(m,\bar m)}$. The result of this simple algebraic computation is well known (e.g., section 3.7 of \cite{Kusuki:2024gtq}): the contribution to the global conformal block due to level-$n$ descendants is simply the $n^\text{th}$ term in the Taylor series of the hypergeometric function in \eqref{eq:Global_s_block}. This allows to decompose the probability $P_m(h_v,h_w;z)$ of populating the double-twist family ${\cal O}_s^{(m,\bar m)}$ into descendant contributions within that family:
\begin{equation}
    \label{eq:Prob_decomp_in_descendants}
    \begin{split}
    P_m(h_v,h_w;z)
    &=\sum_{n=0}^{\infty}P^{(V)}_{m,n}(h_v,h_w;z)
    ~,
    \end{split}
\end{equation}
where the probability to find a descendant of dimension $h_{m,n}$ (in the quantization scheme specified above) is:
\begin{equation}
\label{eq:Prob_descendant_quantization_around_V}
    \begin{split}
    P^{(V)}_{m,n}(h_v,h_w;z) &:=
    \text{probability of populating a level }n\text{ descendant within the conf.\ irrep.\ } h_m
    \\
    &\;= (1-z)^{2h_w}\frac{(2h_v)_m (2h_w)_m}{(2h_v+2h_w+m-1)_m}\frac{(2h_w+m)_n^2}{(2h_v+2h_w+2m)_n}\frac{z^{m+n}}{m!\, n!}~.
    \end{split}
\end{equation}
The superscript ${}^{(V)}$ indicates the choice to expand the OPE around the operator $V$.

We are now in position to compute the expectation value of $L_0$ as a function of the cross ratio $z$:
\begin{equation}
    \label{eq:Expect_L0_sum_of_expvals}
    \mathbb{E}^{(V)}[L_0]=h_v+h_w+\mathbb{E}[m]+\mathbb{E}^{(V)}[n]~.
\end{equation}
Note that the expectation value $\mathbb{E}^{(V)}[n]$ of the descendant dimension depends on the expansion point (here, $V$) and frame, while $\mathbb{E}[m]$ is invariant under $SL(2,\mathbb{R})$ transformations. Explicitly:
\begin{equation}
    \label{eq:Expect_val_of_n_step1}
    \mathbb{E}^{(V)}[n] = \sum_{{m,n}=0}^{\infty}n~P^{(V)}_{m,n}(h_v,h_w;z) = (1-z)^{2h_w} \sum_{m=0}^{\infty}\frac{z^m}{m!}\frac{(2h_v)_m (2h_w)_m}{(2h_v+2h_w+m-1)_m}\sum_{n=1}^{\infty}\frac{(2h_w+m)_n^2}{(n-1)!(2h_v+2h_w+2m)_n}\,z^n~.
\end{equation}
We observe that the sum over $n$ is simply the differential operator $z\partial_z$ acting on the hypergeometric function of the $s$-channel block \eqref{eq:Global_s_block}, namely
\begin{equation}
    \label{eq:ExpVal_n_deriv_2F1}
    \sum_{n=1}^{\infty}\frac{(2h_w+m)_n^2}{(n-1)!(2h_v+2h_w+2m)_n}z^n = z\partial_z\hypergeometricF{2}{1}{2h_w+m,~2h_w+m}{2h_v+2h_w+2m}{z}~,
\end{equation}
for any $z\in[0,1)$ (where the series converges). Also note that $z\partial_z$ is the differential representation of $L_0$ in the complex $z$-plane. Redistributing the derivative, we can rewrite \eqref{eq:Expect_val_of_n_step1} as:
\begin{equation}
    \label{eq:Expect_val_of_n_step2}
    \mathbb{E}^{(V)}[n]=\partial_z \left(z\sum_{m=0}^{\infty}P_{m}(h_v,h_w;z)\right)-\sum_{m=0}^{\infty}\partial_z\bigg((1-z)^{2h_w}z^{m+1}\bigg)\frac{(2h_v)_m(2h_w)_m}{m!(2h_v+2h_w+m-1)_m}\hypergeometricF{2}{1}{2h_w+m,~2h_w+m}{2h_v+2h_w+2m}{z}
\end{equation}
where for the first term we exchanged differentiation and summation signs, relying on the uniform convergence of the sum. The sum in the first term is equal to $1$, by normalization of the distribution $P_m$. After some algebraic manipulations, one can also relate the second term to moments of the primary distribution $P_m$, obtaining:
\begin{equation}
    \label{eq:Expect_val_of_n_result}
    \mathbb{E}^{(V)}[n]=2h_w\,\frac{z}{1-z}-\mathbb{E}[m]~.
\end{equation}
This quantity is non-negative for any $z\in[0,1)$.
Finally, combining \eqref{eq:Expect_val_of_n_result} with \eqref{eq:Expect_L0_sum_of_expvals} the $\mathbb{E}[m]$ term cancels out and we obtain the following global energy expectation value in an OPE expansion centered at $V$:
\begin{equation}
\label{eq:Global_energy_quantiz_around_V_exact}
    \mathbb{E}^{(V)}[L_0] = h_v+h_w+2h_w\,\frac{z}{1-z}~.
\end{equation}
The asymmetry in the dependence on $(h_v,h_w)$ is a consequence of the quantization scheme used to define the descendant states. In this scheme, $W$ is boosted relative to $V$, and hence all the descendants are produced by the action of the boost generator $L_{-1}$ on $W$. The expectation value of the exchanged primary dimension, $\mathbb{E}[m]$, is symmetric in the parameters as it is an invariant within the global family, see \eqref{eq:First_moment}. This also clarifies that $\mathbb{E}[m]$ is a more robust measure for black hole formation and should be related to similarly invariant bulk quantities (in particular the mass $M$).

\subsection{Late-time asymptotics}
\label{app:Asymptotics}

We shall now use the exact results derived above in order to obtain the asymptotic behavior of the mean value and variance of the exchanged quasiprimary dimension as a function of time. We make regular use of the late-time approximation $ e^{t-|b|} \gg 1$, for which the cross ratios take the values indicated in \eqref{eq:Cross_ratios_exact}.

\subsubsection{Average primary dimension}

\begin{figure}
\includegraphics[width=.95\textwidth]{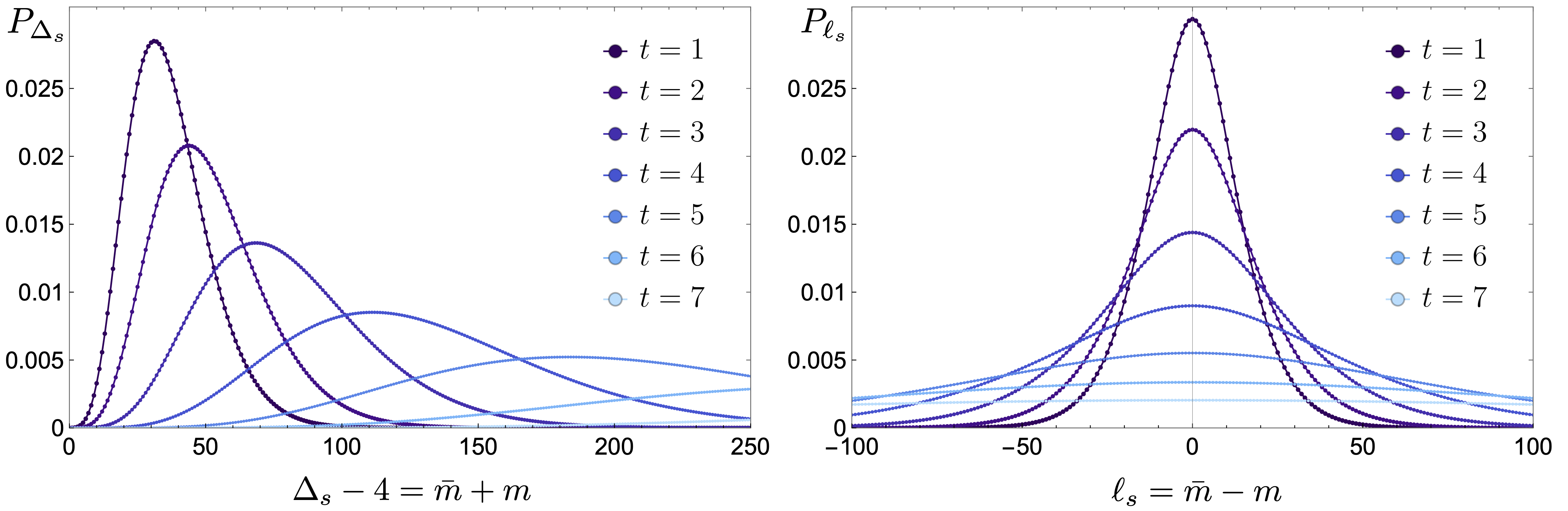}
\caption{Probability distributions for exchange dimension $\Delta_s = \Delta_v+\Delta_w +\bar m + m$ and spin $\ell_s = \ell_v + \ell_w + \bar m - m$ as functions of time. We set $h_v=h_w=\bar h_v=\bar h_w =1$ and $b=0$, thus producing a symmetric distribution of spins with $\mathbb{E}[\ell_s]=0$. A non-zero $\mathbb{E}[\ell_s]$ can be produced either by using spinning external operators, or by setting $b \neq 0$.}
\label{fig:DeltaSpin}
\end{figure}

Near $z\sim1$, the first moment \eqref{eq:First_moment} behaves as:
\begin{equation}
    \label{eq:Expect_m_asympt}
    \mathbb{E}[m]\sim\frac{\Gamma(2h_v+\frac{1}{2}) \Gamma(2h_w+\frac{1}{2})}{\Gamma(2h_v)\Gamma(2h_w)}\,(1-z)^{-1/2}~.
\end{equation}
For $1\ll h_v,h_w \ll c$, we can estimate the Gamma functions using the Stirling approximation. Combining this with \eqref{eq:Cross_ratios_exact}, we reach:
\begin{equation}
    \label{eq:Expect_m_mbar_asympt_t}
    \mathbb{E}[m]\sim \frac{\sqrt{h_v h_w}}{\sin(\delta)}\,e^{(t-b)/2}~,\qquad \mathbb{E}[\bar{m}]\sim \frac{\sqrt{\bar{h}_v \bar{h}_w}}{\sin(\delta)}\,e^{(t+b)/2}~.
\end{equation}
We can now use this to determine the asymptotic behavior of the expected exchanged primary dimension and spin:
\begin{align}
    \mathbb{E}[\Delta_s] \equiv \Delta_v+\Delta_w+\mathbb{E}[\bar{m}+m] &\sim \frac{e^{b/2}\sqrt{(\Delta_v+\ell_v)(\Delta_w+\ell_w)}+e^{-b/2}\sqrt{(\Delta_v-\ell_v)(\Delta_w-\ell_w)}}{2\sin(\delta)}\,e^{t/2}~, \label{eq:Generic_DeltaS_large_t} \\
    \mathbb{E}[\ell_s] \equiv \ell_v+\ell_w+\mathbb{E}[\bar m-m] &\sim \frac{e^{b/2}\sqrt{(\Delta_v+\ell_v)(\Delta_w+\ell_w)}-e^{-b/2}\sqrt{(\Delta_v-\ell_v)(\Delta_w-\ell_w)}}{2\sin(\delta)}\,e^{t/2}~. \label{eq:Generic_SpinS_large_t}
\end{align}
where $\Delta_i=\bar{h}_i+h_i$ and $l_{i}=\bar h_i-h_i$.
Recall from the main text that these expectation values describe properties of Virasoro mean-field wave packets.
Our central proposal is that these should be identified with the mass $M$ and spin $J$ of the conical defect or BTZ black hole created in the bulk collision:
\begin{equation}
    M + \frac{c}{12} := \mathbb{E}[\Delta_s]  \,,\qquad J := \mathbb{E}[\ell_s] \,.
\end{equation}
For illustration, in Fig.\ \ref{fig:DeltaSpin} we show the probability distributions associated with operator dimension and spin. 
Let us also discuss analytically the following two special cases:\\

\noindent{\it $(i)$ Identical operators with no impact parameter.}
Setting $b=0$ in \eqref{eq:Generic_DeltaS_large_t}-\eqref{eq:Generic_SpinS_large_t} given $\Delta_v=\Delta_w\equiv \Delta$ and $\ell_v=\ell_w\equiv \ell$ yields the following simplified expressions for a head-on collision:
\begin{equation}
    \label{eq:Head_on_results}
    M + \frac{c}{12}\sim \frac{\Delta}{\sin(\delta)}~e^{t/2}~,\qquad J\sim \frac{\ell}{\sin(\delta)}~e^{t/2}~.
\end{equation}

\noindent{\it $(ii)$ Scalar operators with arbitrary impact parameter.}
If instead we evaluate \eqref{eq:Generic_DeltaS_large_t}-\eqref{eq:Generic_SpinS_large_t} for 
$\ell_v=\ell_w=0$ taking non-zero impact parameter, we obtain:
\begin{equation}
    \label{eq:Identical_scalars_result}
    M + \frac{c}{12}\sim \frac{\Delta}{\sin(\delta)}\,\cosh(\frac{b}{2})\,e^{t/2} ~,\qquad J\sim \frac{\Delta}{\sin(\delta)}\,\sinh(\frac{b}{2})\,e^{t/2} ~,
\end{equation}
where $\Delta\equiv \sqrt{\Delta_v \Delta_w}$.\\

In each case, we obtain the black hole formation timescale $t_{\text{BH}} \sim 2 t_*$ by evaluating the BTZ extremality condition $M = |J|$, as shown in the main text \cite{BTZ1,BTZ2,Keller:2014xba}.

\subsubsection{Second moment and variance}

In order to justify the claim that the distribution of exchanged primaries forms a localized wave packet, it is important that the statistical variance does not grow faster than the mean. In this section we compute the second moment and the variance:
\begin{equation}
    \label{eq:Variance_def}
    {\rm Var}[m]:= \mathbb{E}[m^2]-\mathbb{E}[m]^2~.
\end{equation}
Evaluating the exact second moment \eqref{eq:Second_moment} on the late-time cross ratios \eqref{eq:Cross_ratios_exact}, we find:
\begin{equation}
    \mathbb{E}[m^2] \sim \frac{h_v h_w}{\sin^2(\delta)} \, e^{t-b}\,,\qquad
    \mathbb{E}[\bar{m}^2] \sim \frac{\bar h_v \bar h_w}{\sin^2(\delta)} \, e^{t+b} \,.
\end{equation}
This shows that $\mathbb{E}[m^2]$ is of the same order as $\mathbb{E}[m]^2$, i.e., their ratio tends to a constant:
\begin{equation}
    \frac{\mathbb{E}[m^2]}{\mathbb{E}[m]^2} \sim \frac{\Gamma(2h_v)\Gamma(2h_v+1) \Gamma(2h_w)\Gamma(2h_w+1)}{\Gamma(2h_v+\frac{1}{2})^2 \Gamma(2h_w+\frac{1}{2})^2} + \mathit{O}\big( e^{-t/2} \big)\,.
\end{equation}
Correspondingly, the variance does not vanish at order $e^t$. Computing the first and second moment to sufficiently high order, we find for $z \rightarrow 1^-$:
\begin{equation}
    \label{eq:Var_asymptotic}
    {\rm Var}[m]\sim \left[ 4h_vh_w -\left(\frac{\Gamma(2h_v+\frac{1}{2})\Gamma(2h_w+\frac{1}{2})}{\Gamma(2h_v)\Gamma(2h_w)}\right)^2 \right] (1-z)^{-1}  + \mathit{O}\left((1-z)^0\right)~,
\end{equation}

\subsubsection{Higher moments}

For completeness, we note that the asymptotic behavior of the expectation value of the $p$-th falling factorial near $z\sim 1$ is
\begin{equation}
    \label{eq:Falling_fact_asympt}
    \mathbb{E}[m^{\underline{p}}] = S_p\left(2h_v,2h_w,2h_w+2h_w;\frac{z}{z-1}\right)\sim \frac{\Gamma(2h_v+\frac{p}{2})\Gamma(2h_w+\frac{p}{2})}{\Gamma(2h_v)\Gamma(2h_w)}\,(1-z)^{-p/2}+\mathit{O}\left((1-z)^{-(p-1)/2}\right)~.
\end{equation}
Using \eqref{eq:Moments_from_falling_factorials}, one may translate this into the asymptotic behavior of the moments $\mathbb{E}[m^k]$.  
Given that the Stirling numbers of the second kind do not depend on $z$, and that $\mathcal{S}(k,k)=1$ for all $k\geq 0$, it follows that the leading behavior of $\mathbb{E}[m^k]$ is equal to that of $\mathbb{E}[m^{\underline{k}}]$ as $z\rightarrow 1^-$.
We conclude that the $k$-th moment of the primary distribution $\mathbb{E}[m^k]$ behaves as $(1-z)^{-k/2}$, times a prefactor that depends on $h_v,\,h_w$ and $k$. Consequently, \begin{equation}
\frac{\mathbb{E}[m^k]}{\mathbb{E}[m]^k} \sim \text{const.}+ {\cal O}\left( 1-z \right)\,.
\end{equation}

\newpage
\twocolumngrid 
\bibliographystyle{apsrev}
\bibliography{refs.bib}

\end{document}